\documentstyle[aps,epsfig]{revtex}

\begin{document}
\title{Metal-insulator transition in the Hubbard model: a simple description
including the Kondo effect}
\author{V. Yu. Irkhin$^{*}$ and A. V. Zarubin}
\address{Institute of Metal Physics, S.Kovalevskay str. 18, 620219 Ekaterinburg,\\
Russia}
\maketitle

\begin{abstract}
The electron spectrum structure in the half-filled Hubbard model is
considered in terms of the one-particle Green's functions within
many-electron representation. A simple analytical generalization of the
single-site Hubbard-III approximation is obtained, which takes into account
the Fermi excitations (Kondo terms). The problem of the metal-insulator
transition is investigated. The occurrence of a three-peak density-of-states
structure including the ``Kondo'' peak at the Fermi level is discussed. A
comparison with large-$d$ calculations is performed.
\end{abstract}


\section{Introduction}

\label{sec:intro}

The problem of strong correlations in many-electron systems is one of the
most important in the solid state theory. The simplest model to describe
correlation effects is the Hubbard model \cite{Hubbard-I:1963} which
includes the on-site Coulomb interaction. One of most interesting phenomena
is the correlation-driven metal-insulator transition (MIT), which takes
place in a number of transition metal compounds. A simple description of MIT
was given by Hubbard \cite{Hubbard-III:1964} who started from the
atomic-level picture and proposed a simple interpolation self-consistent
scheme.

Since the Hubbard works of 60's, a great progress has been achieved in
understanding electronic structure of highly-correlated systems. Last time,
the role of the Kondo effect has been discussed within the large-$d$
approach ($d$ is space dimensionality) which reduces the original periodic
Hubbard model to an effective Anderson impurity model \cite
{larged,larged1,Rozenberg:1994}. Such an approach (dynamical mean-field
theory, DMFT) turned out to be rather successful. The corresponding density
of states (DOS) has three-peak rather than two-peak structure: an additional
``Kondo'' quasiparticle resonance at the Fermi level occurs owing to
scattering by the local moment. The spectrum structure in large-$d$
approaches is confirmed by the quantum Monte-Carlo (QMC) calculations (see,
e.g., \cite{Schlipf:1999}) and some spectroscopic experimental results.
Unfortunately, there exist some difficulties in numerical calculations
within QMC and large-$d$ approaches, so that one needs often to introduce
rather high temperatures to resolve these problems. The three-peak structure
is not reproduced by most preceding analytical approaches, in particular, by
the single-site Hubbard-III approximation \cite{Hubbard-III:1964}, the
reason being in that they do not take into account contributions of
Fermi-like excitations in a proper way. Thus these approaches do not
describe the Brinkman-Rice effective-mass enhancement which is important
from the experimental point of view. Recently, an attempt has been made to
improve the Hubbard-III approximation by calculating corrections owing to
correlation effects \cite{Luo:2000}; however, the results remained
qualitatively unchanged.

A detailed analysis of Hubbard-III-like approximations was performed in
Refs.\ \cite{Anokhin:1991,Anokhin:1991a} within the large-$z$ expansion, $z$
being the nearest-neighbor number. In the zero order this approach reduces
to the simplest Hubbard-I approximation \cite{Hubbard-I:1963}. General
expressions for $1/z$-corrections in the limit $U\rightarrow \infty $ were
obtained in Ref.\ \cite{Anokhin:1991}. The problem of MIT within this
approach was treated in Ref.\ \cite{Anokhin:1991a}. Unfortunately, only a
classical approximation (the large-$S$ limit of the $s$-$d$ model which
generalizes the Hubbard model) was considered, and the terms with the
one-particle occupation numbers, which just describe the Kondo effect in
narrow bands \cite{Irkhin:1999}, were neglected.

In the present paper we present a treatment that is based on the method of
equations of motion for the many-electron Hubbard operators \cite
{Hubbard-IV:1965,Irkhin:1994} and is a much more simple than the large-$d$
approach. In Sec.\ \ref{sec:ds}, the decoupling scheme with account of the
Fermi excitations is developed. In Sec.\ \ref{sec:rd}, we present the
results of numerical calculations and carry out a comparison with previous
works.

\section{The decoupling scheme}

\label{sec:ds}

We start from the Hubbard model with the electron concentration $n=1$ (the
half-filled case). The corresponding Hamiltonian reads
\begin{equation}
{\cal H}=\sum_{{\bf k}\sigma }t_{{\bf k}}c_{{\bf k}\sigma }^{\dag } c_{{\bf k%
}\sigma }+U\sum_in_{i\uparrow }n_{i\downarrow },  \label{eq:HHM}
\end{equation}
where $n_{i\sigma }=c_{i\sigma }^{\dag }c_{i\sigma }$, $c_{i\sigma }^{\dag }$
and $c_{{\bf k}\sigma }^{\dag }$ are the one-electron operators in the
Wannier and quasimomentum representation. We pass to the Hubbard $X$%
-operators
\begin{equation}
X_i^{\alpha \beta }=|i\alpha \rangle \langle i\beta |,\quad X_i^{\alpha
\beta }X_i^{\gamma \varepsilon }=\delta _{\beta \gamma }X_i^{\alpha
\varepsilon },\quad \sum_\alpha X_i^{\alpha \alpha }=1,  \label{eq:comm}
\end{equation}
so that
\begin{equation}
c_{i\sigma }^{\dag }=\sum_{\alpha ,\beta }\langle i\alpha |c_{i\sigma
}^{\dagger }|i\beta \rangle X_i^{\alpha \beta }=X_i^{\sigma 0}+\sigma
X_i^{2-\sigma }.  \label{eq:c2X}
\end{equation}
Then the interaction Hamiltonian takes a diagonal form and we obtain
\begin{equation}
{\cal H}=\sum_{{\bf k}\sigma }t_{{\bf k}}(X_{-{\bf k}}^{\sigma 0} +\sigma
X_{-{\bf k}}^{2-\sigma }) (X_{{\bf k}}^{0\sigma }+\sigma X_{{\bf k}%
}^{-\sigma 2})+U\sum_iX_i^{22}.  \label{eq:HHM:X}
\end{equation}
Using (\ref{eq:c2X}) we have for the one-electron anticommutator retarded
Green's function
\begin{equation}
G_{{\bf k}\sigma }(E)=\langle \!\langle c_{{\bf k}\sigma }|c_{{\bf k}\sigma
}^{\dagger }\rangle \!\rangle _E=\langle \!\langle X_{{\bf k}}^{0\sigma }|c_{%
{\bf k}\sigma }^{\dagger }\rangle \!\rangle _E+\sigma \langle \!\langle X_{%
{\bf k}}^{-\sigma 2}|c_{{\bf k}\sigma }^{\dagger }\rangle \!\rangle _E.
\label{eq:GF:0}
\end{equation}
The energy $E$ is supposed to be referred to the chemical potential which
equals $U/2$ in our case. We write down the system of equation of motion
\[
E\langle \!\langle A|B\rangle \!\rangle _E=\langle \{A,B\}\rangle +\langle
\!\langle [A,{\cal H}]|B\rangle \!\rangle _E
\]
for the pair of the Green's functions in the right-hand side of (\ref
{eq:GF:0}). Using (\ref{eq:comm}) we obtain in the non-magnetic case
\begin{eqnarray}
(E+U/2)\langle \!\langle X_{{\bf k}}^{0\sigma }|c_{{\bf k}\sigma }^{\dagger
}\rangle \!\rangle _E &=&\frac 12(1+t_{{\bf k}}\langle \!\langle c_{{\bf k}%
\sigma }|c_{{\bf k}\sigma }^{\dagger }\rangle \!\rangle _E)  \nonumber \\
&&+\sum_{{\bf q}}t_{{\bf q}}\langle \!\langle (\delta (X_{{\bf k-q}%
}^{00})+\delta (X_{{\bf k}-{\bf q}}^{\sigma \sigma }))c_{{\bf q}\sigma }|c_{%
{\bf k}\sigma }^{\dagger }\rangle \!\rangle _E  \nonumber \\
&&+\sum_{{\bf q}}t_{{\bf q}}\langle \!\langle X_{{\bf k}-{\bf q}}^{-\sigma
\sigma }c_{{\bf q}-\sigma }+\sigma c_{{\bf q}-\sigma }^{\dagger }X_{{\bf k+q}%
}^{02}|c_{{\bf k}\sigma }^{\dagger }\rangle \!\rangle _E,  \label{eq:EM:1} \\
\sigma (E-U/2)\langle \!\langle X_{{\bf k}}^{-\sigma 2}|c_{{\bf k}\sigma
}^{\dagger }\rangle \!\rangle _E &=&\frac 12(1+t_{{\bf k}}\langle \!\langle
c_{{\bf k}\sigma }|c_{{\bf k}\sigma }^{\dagger }\rangle \!\rangle _E)
\nonumber \\
&&+\sum_{{\bf q}}t_{{\bf q}}\langle \!\langle (\delta (X_{{\bf k-q}%
}^{-\sigma -\sigma })+\delta (X_{{\bf k-q}}^{22}))c_{{\bf q}\sigma }|c_{{\bf %
k}\sigma }^{\dagger }\rangle \!\rangle _E  \nonumber \\
&&-\sum_{{\bf q}}t_{{\bf q}}\langle \!\langle X_{{\bf k-q}}^{-\sigma \sigma
}c_{{\bf q}-\sigma }+\sigma c_{{\bf q}-\sigma }^{\dagger }X_{{\bf k+q}%
}^{02}|c_{{\bf k}\sigma }^{\dagger }\rangle \!\rangle _E,  \label{eq:EM:2}
\end{eqnarray}
where $\delta A=A-\langle A\rangle $ is the fluctuation of the operator.
Solving the system (\ref{eq:EM:1}), (\ref{eq:EM:2}) we derive
\begin{equation}
G_{{\bf k}\sigma }(E)=G_{{\bf k}}^0(E)\left( 1-\frac UE\Gamma _{{\bf k}%
\sigma }(E)\right) ,  \label{eq:GF:1}
\end{equation}
\begin{equation}
\Gamma _{{\bf k}\sigma }(E)=\sum_{{\bf q}}t_{{\bf q}}\langle \!\langle
\delta (X_{{\bf k-q}}^{00}+X_{{\bf k}-{\bf q}}^{\sigma \sigma })c_{{\bf q}%
\sigma }+X_{{\bf k}-{\bf q}}^{-\sigma \sigma }c_{{\bf q}-\sigma }+\sigma c_{%
{\bf q}-\sigma }^{\dagger }X_{{\bf k+q}}^{02}|c_{{\bf k}\sigma }^{\dagger
}\rangle \!\rangle _E.  \label{eq:GF:1a}
\end{equation}
Here
\begin{equation}
G_{{\bf k}}^0(E)=\frac 1{F_0(E)-t_{{\bf k}}},\quad F_0(E)=E-\frac{U^2}{4E}
\label{eq:GF:HI}
\end{equation}
is the Green's function of the Hubbard-I approximation (which plays the role
of a mean-filed approximation for our problem), $F_0(E)$ being the
corresponding inverse locator. The Hubbard-I spectrum contains two
correlation subbands defined by the poles of (\ref{eq:GF:HI})
\[
E_{{\bf k}1,2}=\frac 12(t_{{\bf k}}\pm \varepsilon _{{\bf k}}),\quad
\varepsilon _{{\bf k}}=\sqrt{U^2+t_{{\bf k}}^2}.
\]
The Green's function $\Gamma _{{\bf k}\sigma }(E)$ describes fluctuation
corrections. The corresponding collective excitations are described by spin
and charge operators.
\begin{eqnarray*}
S_{{\bf q}}^\sigma &=&X_{{\bf q}}^{\sigma -\sigma },\quad S_{{\bf q}}^z=%
\frac 12(X_{{\bf q}}^{++}-X_{{\bf q}}^{--}), \\
\rho _{{\bf q}}^{+} &=&X_{{\bf q}}^{20},\quad \rho _{{\bf q}}^z=\frac 12(X_{%
{\bf q}}^{22}-X_{{\bf q}}^{00}).
\end{eqnarray*}

Further we write down the system of equations for the fluctuation Green's
functions and perform the decouplings which correspond to the first order in
the formal parameter $1/z$ (strictly speaking, this expansion is justified
in the case of long-range electron hopping). For the half-filled band, we
have to take into account particle and hole excitations in an equal way.
However, decouplings can violate the particle-hole symmetry. To preserve
this symmetry, we make an identical transformation by taking in (\ref
{eq:GF:1a}) the Green's functions with symmetrized operator products, e.g.,
\[
\langle \!\langle X_{{\bf k}-{\bf q}}^{-\sigma \sigma }c_{{\bf q}-\sigma
}|c_{{\bf k}\sigma }^{\dagger }\rangle \!\rangle _E\rightarrow \frac 12%
\langle \!\langle \{X_{{\bf k}-{\bf q}}^{-\sigma \sigma },c_{{\bf q}-\sigma
}\}|c_{{\bf k}\sigma }^{\dagger }\rangle \!\rangle _E=\frac 12\langle
\!\langle X_{{\bf k}-{\bf q}}^{-\sigma \sigma }c_{{\bf q}-\sigma }+c_{{\bf q}%
-\sigma }X_{{\bf k}-{\bf q}}^{-\sigma \sigma }|c_{{\bf k}\sigma }^{\dagger
}\rangle \!\rangle _E.
\]
We also use in the equations of motion the Hamiltonian in the symmetrized
form,
\[
t_{{\bf k}}c_{{\bf k}\sigma }^{\dag }c_{{\bf k}\sigma }\rightarrow \frac 12%
t_{{\bf k}}(c_{{\bf k}\sigma }^{\dag }c_{{\bf k}\sigma }-c_{{\bf k}\sigma
}c_{{\bf k}\sigma }^{\dag }).
\]
Then we obtain for the transverse spin fluctuation contribution
\begin{eqnarray*}
(E^2-U^2/4-Et_{{\bf q}})\langle \!\langle \frac 12\{X_{{\bf k}-{\bf q}%
}^{-\sigma \sigma },c_{{\bf q}-\sigma }\}|c_{{\bf k}\sigma }^{\dagger
}\rangle \!\rangle _E &=&(E+U/2)(t_{{\bf q}}-t_{{\bf k}})(f_{{\bf q}}-\frac 1%
2)G_{{\bf k}\sigma }(E)-U(t_{{\bf k}}\langle X_{{\bf k}-{\bf q}}^{-\sigma
\sigma }X_{{\bf -k+q}}^{\sigma -\sigma }\rangle \\
&&+(t_{{\bf q}}-t_{{\bf k}})\langle \frac 12[X_{-{\bf q}}^{-\sigma 0},X_{%
{\bf q}}^{0-\sigma }]-\sigma X_{-{\bf q}}^{2\sigma }X_{{\bf q}}^{0-\sigma
}\rangle G_{{\bf k}\sigma }(E) \\
&&-(E+U/2)(f_{{\bf q}}-\frac 12)+U\langle \frac 12[X_{-{\bf q}}^{-\sigma
0},X_{{\bf q}}^{0-\sigma }] \\
&&-\sigma X_{-{\bf q}}^{2\sigma }X_{{\bf q}}^{0-\sigma }-X_{{\bf k}-{\bf q}%
}^{-\sigma \sigma }X_{-{\bf k+q}}^{\sigma -\sigma }\rangle .
\end{eqnarray*}
Note that, as well as the standard Kondo terms, the ``many-electron'' terms
come from the spin-flip processes, but not from longitudinal spin
fluctuations. For the ``transverse'' charge contribution we have
\begin{eqnarray*}
\sigma (E^2-U^2/4-Et_{{\bf q}})\langle \!\langle \frac 12\{c_{{\bf q}-\sigma
}^{\dagger }X_{{\bf k+q}}^{02}\}|c_{{\bf k}\sigma }^{\dagger }\rangle
\!\rangle _E &=&(E+U/2)(t_{{\bf q}}+t_{{\bf k}})(f_{{\bf q}}-\frac 12)G_{%
{\bf k}\sigma }(E) \\
&&+U(t_{{\bf k}}\langle X_{{\bf -k}-{\bf q}}^{20}X_{{\bf k+q}}^{02}\rangle
+(t_{{\bf q}}+t_{{\bf k}})\langle \sigma X_{-{\bf q}}^{2\sigma }X_{{\bf q}%
}^{0-\sigma } \\
&&+\frac 12[X_{-{\bf q}}^{2\sigma },X_{{\bf q}}^{\sigma 2}]\rangle )G_{{\bf k%
}\sigma }(E) \\
&&+(E+U/2)(f_{{\bf q}}-\frac 12)-U\langle \frac 12[X_{-{\bf q}}^{2\sigma
},X_{{\bf q}}^{\sigma 2}] \\
&&-\sigma X_{-{\bf q}}^{2\sigma }X_{{\bf q}}^{0-\sigma }-X_{-{\bf k-q}%
}^{20}X_{{\bf k+q}}^{02}\rangle .
\end{eqnarray*}
A symmetry of spin and charge degrees of freedom occurs for a symmetric
conduction band.

Further we neglect ${\bf q}$-dependence of spin and charge correlations
functions and replace them by single-site averages, so that
\begin{eqnarray}
\langle S_{-{\bf q}}^\sigma S_{{\bf q}}^{-\sigma }\rangle  &=&2\langle S_{%
{\bf -q}}^zS_{{\bf q}}^z\rangle =\langle X^{\sigma \sigma }\rangle ,
\nonumber \\
\langle \rho _{-{\bf q}}^\sigma \rho _{{\bf q}}^{-\sigma }\rangle
&=&2\langle \rho _{-{\bf q}}^z\rho _{{\bf q}}^z\rangle =\langle
X^{22}\rangle =\langle X^{00}\rangle .  \label{ons}
\end{eqnarray}
Such an approximation is made (although as a rule implicitly) in practically
all works on the MIT problem. This corresponds to neglecting dynamics of
low-energy Bose excitations and may be justified not only in
high-temperature limit, but also within the $1/z$-expansion. For the
local-spin subsystem, this approximation is in spirit of the mean-field
theory. Main part of charge dynamics (the Hubbard splitting $U$) is also
already taken into account in the zero-order (Hubbard-I) approximation. A
consistent consideration of dynamics can be made in higher orders in $1/z$.
This may lead to a change in details of the MIT picture. However, this
problem is rather difficult.

Taking into account (\ref{ons}) we can use the sum rule in (\ref{eq:comm})
to obtain
\begin{eqnarray}
G_{{\bf k}}(E)=\frac{a(E)}{b(E)-a(E)t_{{\bf k}}}=\frac 1{F(E)-t_{{\bf k}}}%
,\quad F(E)=\frac{b(E)}{a(E)},  \label{eq:GF:2}
\end{eqnarray}
\begin{mathletters}
\label{eq:GF:2ab}
\begin{eqnarray}
a(E) &=&1+\frac 34\frac{U^2}{E^2}\sum_{{\bf q}}t_{{\bf q}}\frac 1{F_0(E)-t_{%
{\bf q}}}+\frac{2U}E\sum_{{\bf q}}t_{{\bf q}}\frac{f_{{\bf q}}}{F_0(E)-t_{%
{\bf q}}},  \label{eq:GF:2a} \\
b(E) &=&F_0(E)+\frac{2U}E\sum_{{\bf q}}t_{{\bf q}}^2\frac{f_{{\bf q}}}{%
F_0(E)-t_{{\bf q}}}.  \label{eq:GF:2b}
\end{eqnarray}
We have substituted here the one-particle correlation functions in the
Hubbard-I approximation,
\end{mathletters}
\begin{eqnarray}
\langle c_{{\bf q}\sigma }^{\dagger }X_{{\bf q}}^{0\sigma }\rangle &=&\frac 1%
{2\varepsilon _{{\bf q}}}[(E_{{\bf q}1}-U)f(E_{{\bf q}1})-(E_{{\bf q}%
2}-U)f(E_{{\bf q}2})],  \nonumber \\
\langle c_{{\bf q}\sigma }^{\dagger }X_{{\bf q}}^{-\sigma 2}\rangle &=&\frac{%
-\sigma }{2\varepsilon _{{\bf q}}}[E_{{\bf q}1}f(E_{{\bf q}1})-E_{{\bf q}%
2}f(E_{{\bf q}2})],  \nonumber
\end{eqnarray}
\[
f_{{\bf q}}\equiv \langle c_{{\bf q}\sigma }^{\dagger }c_{{\bf q}\sigma
}\rangle =\frac 1{\varepsilon _{{\bf q}}}[(E_{{\bf q}1}-U/2)f(E_{{\bf q}%
1})-(E_{{\bf q}2}-U/2)f(E_{{\bf q}2})].
\]
Due to the symmetry of the bare band, we have
\[
\sum_{{\bf q}}f(E_{{\bf q,}2})\Phi (t_{{\bf q}})=\sum_{{\bf q}}[1-f(E_{{\bf q%
}1})]\Phi (-t_{{\bf q}}).
\]
As a result of our way of decoupling, we have in the sums $f(E_{{\bf q}%
1})\rightarrow f(E_{{\bf q}1})-1/2$, and the DOS of the interacting system
remains symmetric.

\begin{figure}[tbp]
\epsfig{file=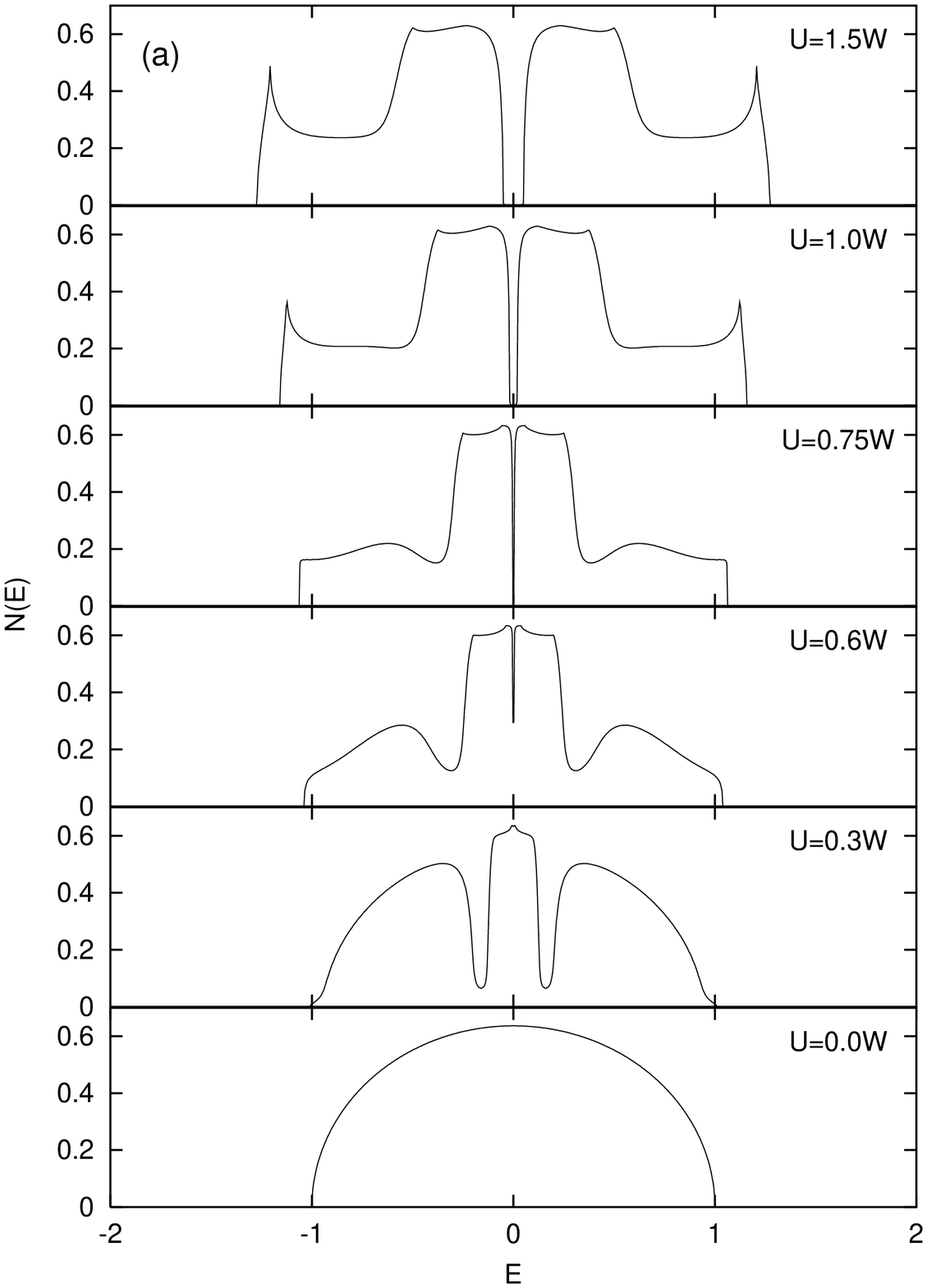,width=8cm} \epsfig{file=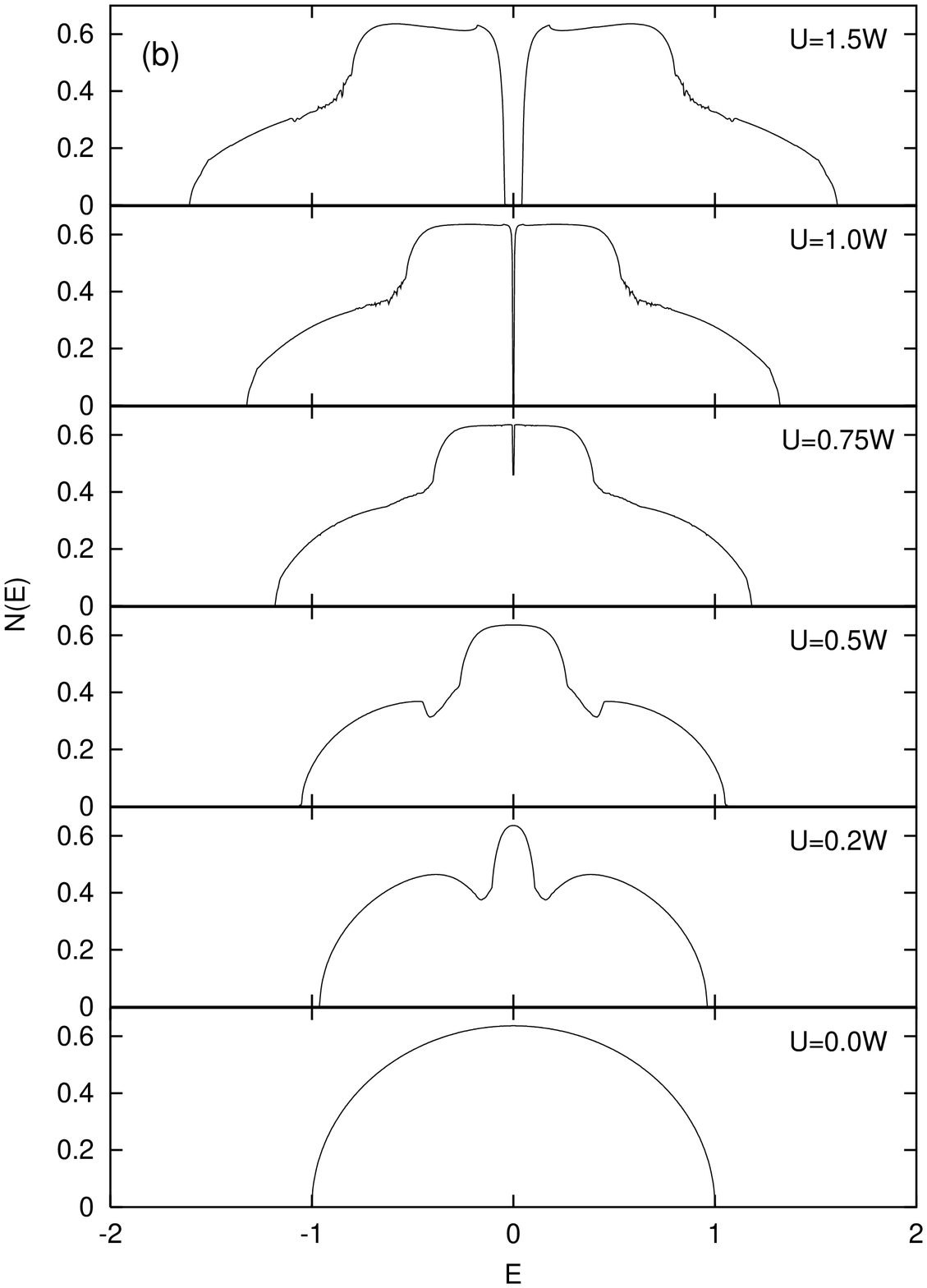,width=8cm}
\caption{Density of states for the semielliptic DOS (a)
approximation (\ref {eq:GF:2ab}) (b) SC approximation
(\ref{eq:GF:2ab:s-c}).}
\label{fig:1}
\end{figure}

\begin{figure}[tbp]
\epsfig{file=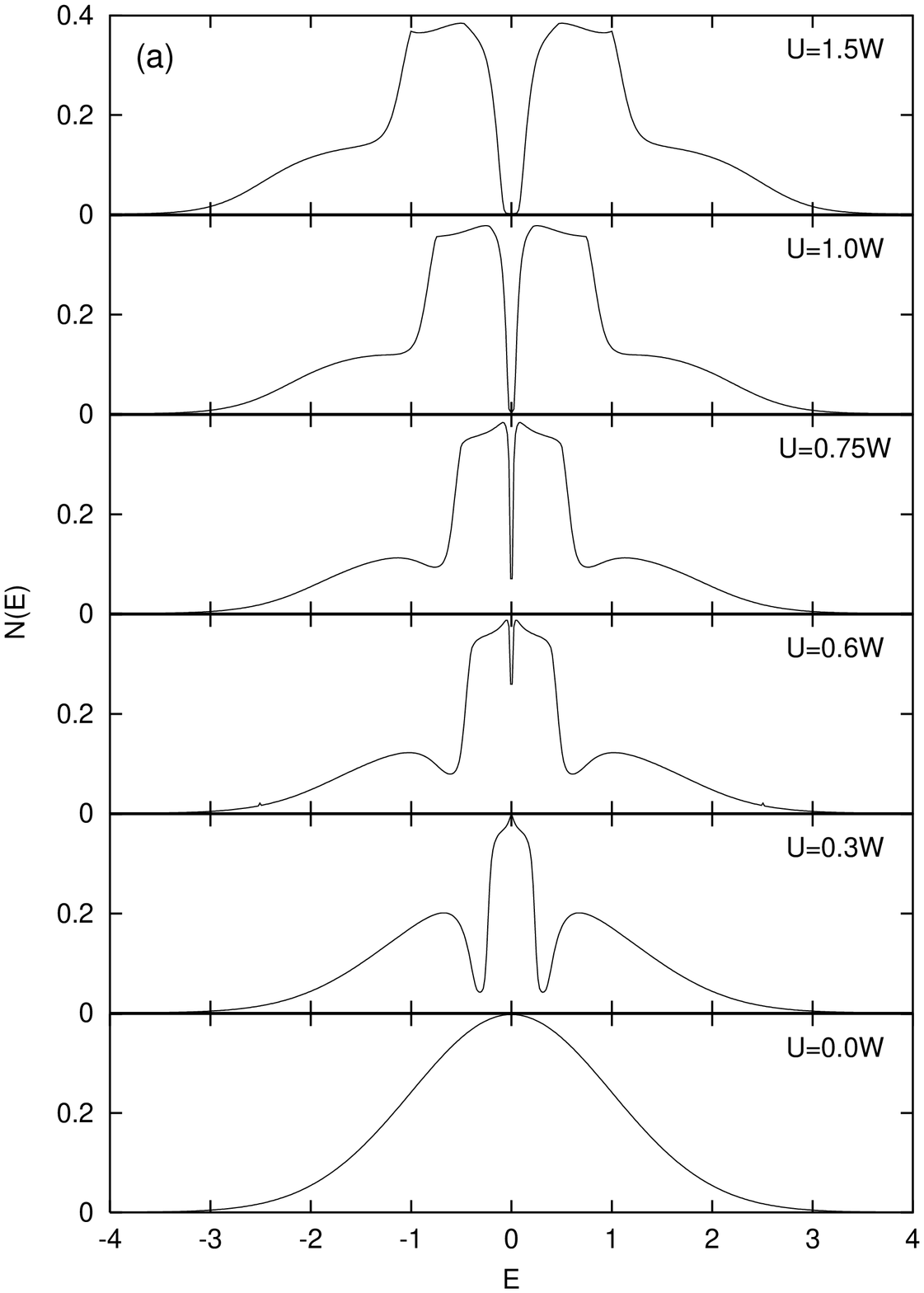,width=8cm} \epsfig{file=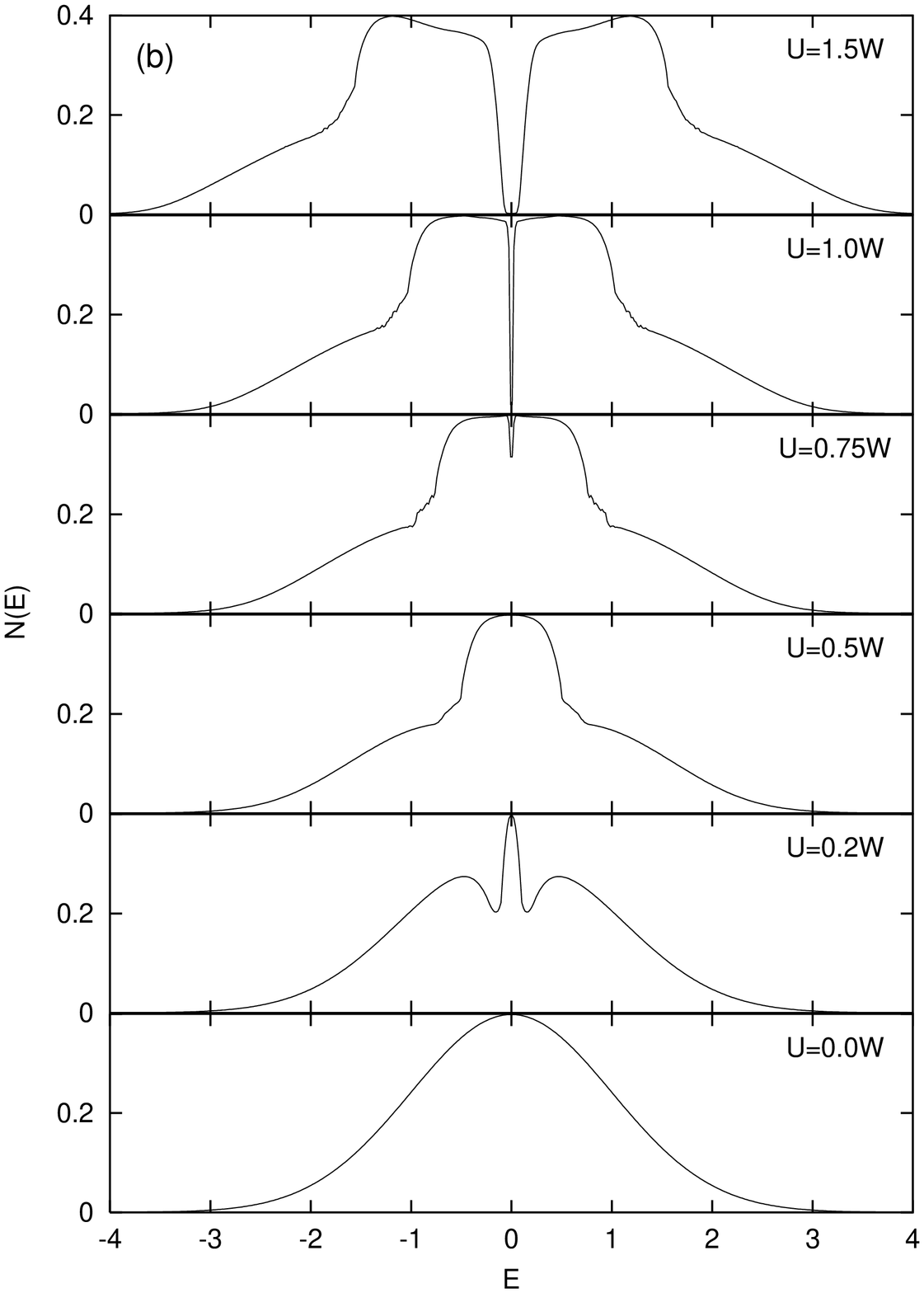,width=8cm}
\caption{Density of states for the Gussian DOS (a) approximation
(\ref {eq:GF:2ab}) (b) SC approximation (\ref{eq:GF:2ab:s-c}).}
\label{fig:2}
\end{figure}

\begin{figure}[tbp]
\epsfig{file=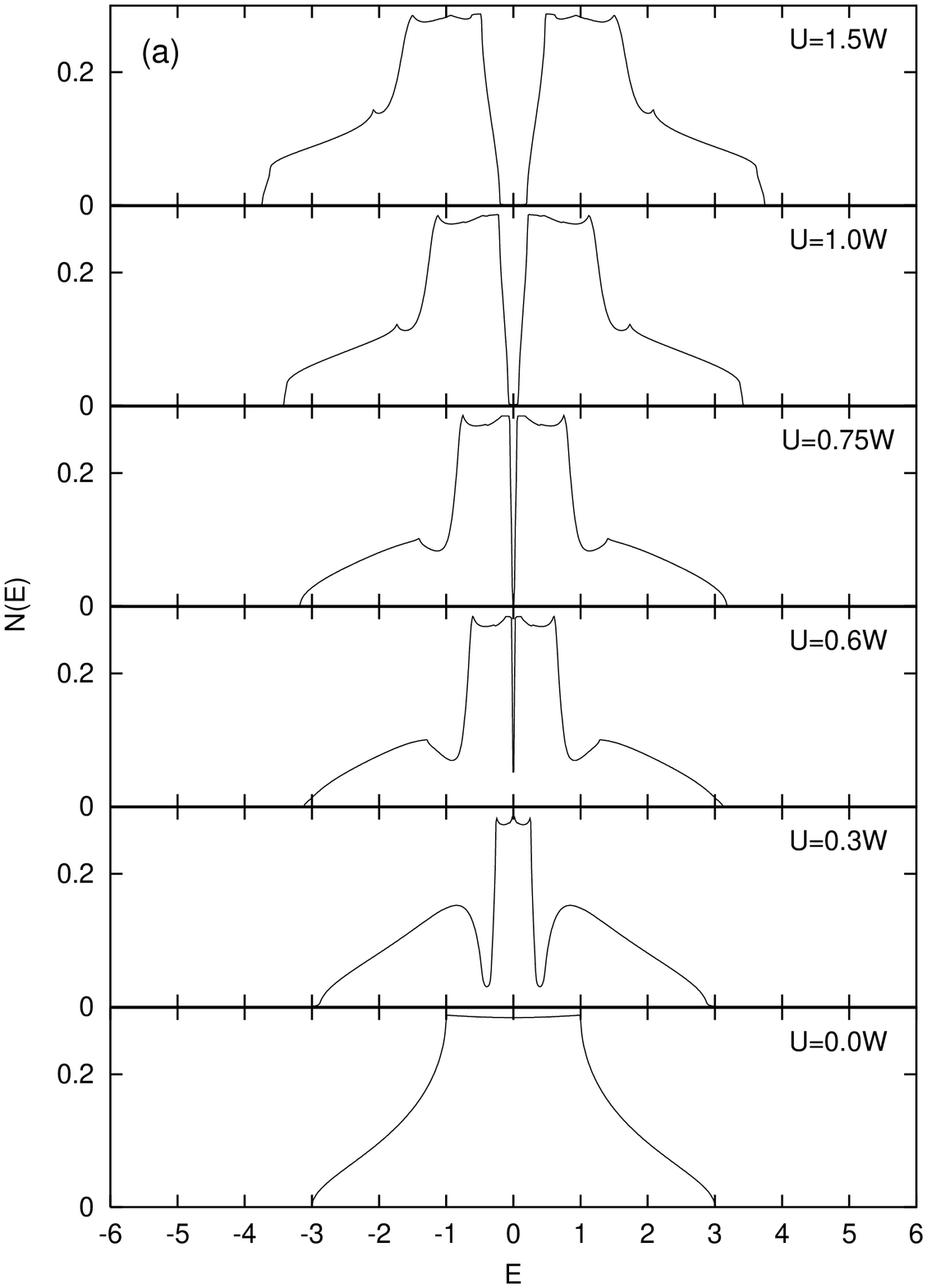,width=8cm} \epsfig{file=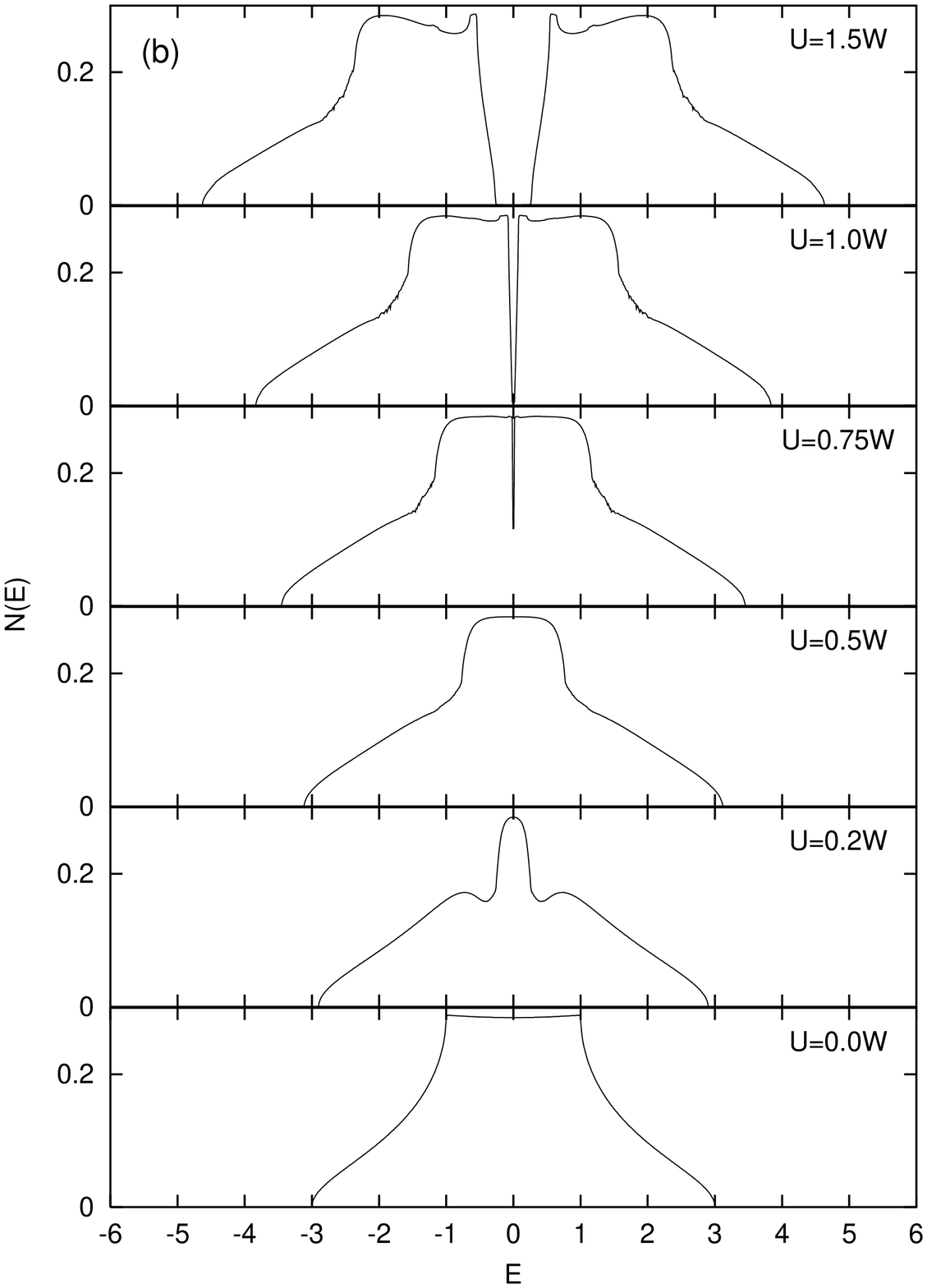,width=8cm}
\caption{Density of states for the simple cubic lattice (a) approximation (%
\ref{eq:GF:2ab}) (b) SC approximation (\ref{eq:GF:2ab:s-c}).}
\label{fig:3}
\end{figure}

\begin{figure}[tbp]
\epsfig{file=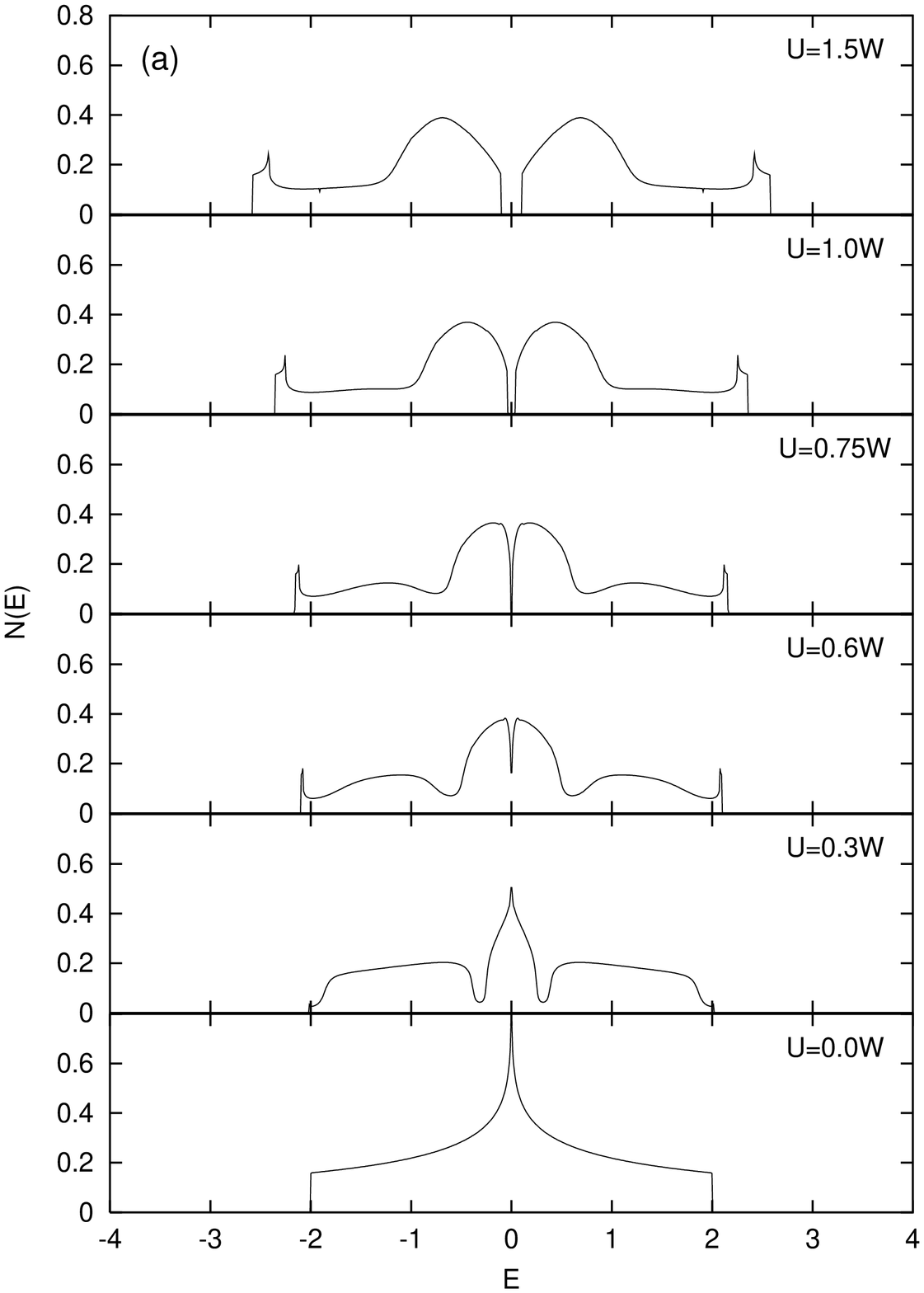,width=8cm} \epsfig{file=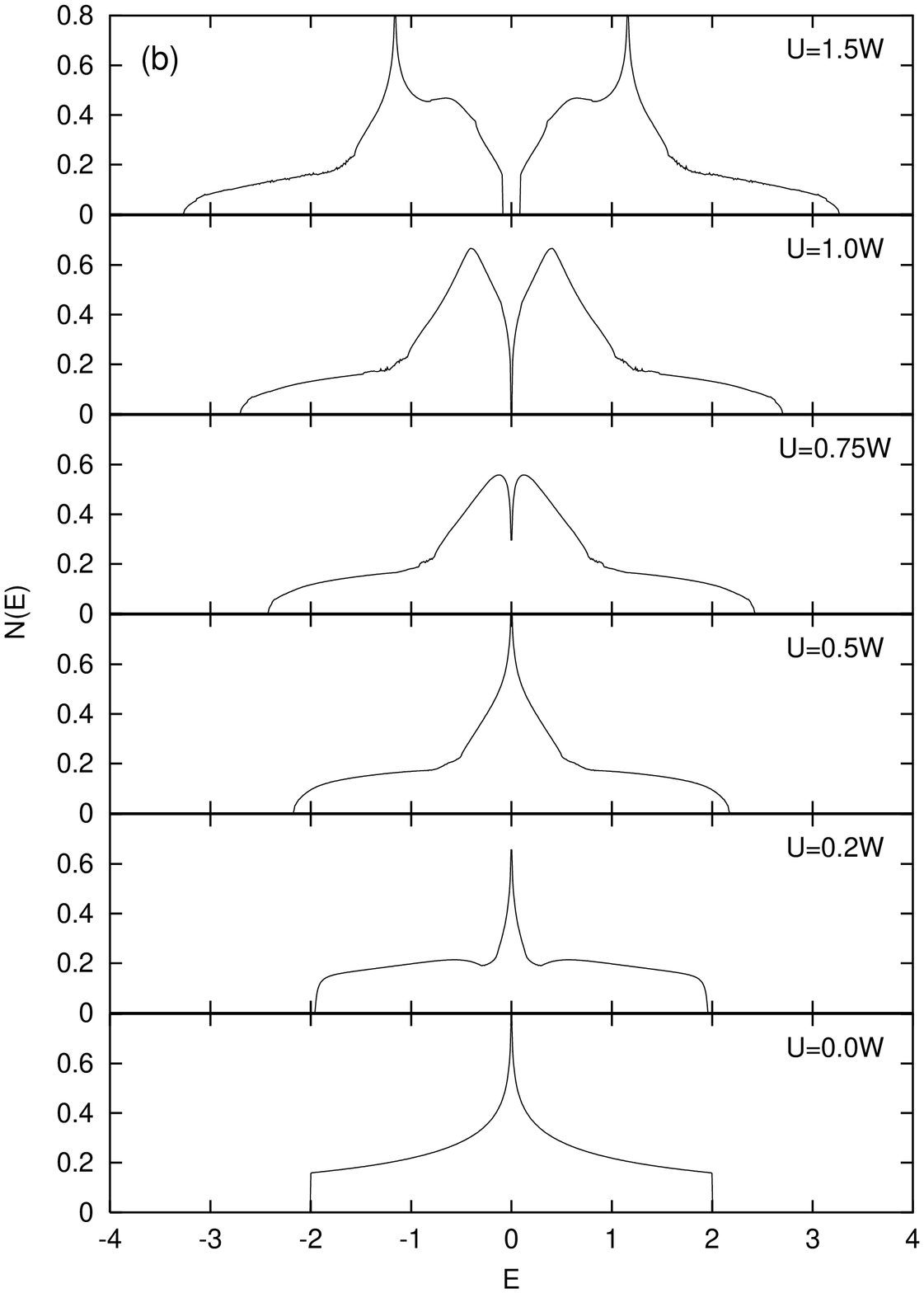,width=8cm}
\caption{Density of states for the square lattice (a)
approximation (\ref {eq:GF:2ab}) (b) SC approximation
(\ref{eq:GF:2ab:s-c}).}
\label{fig:4}
\end{figure}

To obtain the self-consistent (SC) approximation we have to replace in (\ref
{eq:GF:HI}) the Hubbard-I Green's functions by the exact ones,
\[
G_{{\bf q}}^0(E)=\frac 1{F_0(E)-t_{{\bf q}}}\rightarrow G_{{\bf q}}(E)=\frac
1{F(E)-t_{{\bf q}}},
\]
and the Fermi functions $f_{{\bf q}}$ by the exact occupation numbers $n_{%
{\bf q}}$, according to the spectral representation,
\[
n_{{\bf q}}=-\frac 1\pi \int dEf(E){\rm Im}G_{{\bf q}}(E).
\]
Then we have the SC equation for the one-electron Green's function in the
form (\ref{eq:GF:2}) with
\begin{mathletters}
\label{eq:GF:2ab:s-c}
\begin{eqnarray}
a(E) &=&1+\frac 34\frac{U^2}{E^2}\sum_{{\bf q}}t_{{\bf q}}G_{{\bf q}}(E)+%
\frac{2U}E\sum_{{\bf q}}t_{{\bf q}}G_{{\bf q}}(E)n_{{\bf q}},
\label{eq:GF:2a:s-c} \\
b(E) &=&F_0(E)+\frac{2U}E\sum_{{\bf q}}t_{{\bf q}}^2G_{{\bf q}}(E)n_{{\bf q}%
}.  \label{eq:GF:2b:s-c}
\end{eqnarray}

Unlike the simplest self-consistency scheme considered in Ref.\ \cite
{Anokhin:1991} [see eq.~(32) of that paper], the approximation (\ref
{eq:GF:2ab:s-c}), as well as the standard Hubbard-III approximation, does
not result in a violation of analytical properties.

\section{Results and discussion}

\label{sec:rd}

To investigate the MIT problem, we calculate the single-particle density of
states
\end{mathletters}
\[
N(E)=-\frac 1\pi {\rm Im}\sum_{{\bf k}}G_{{\bf k}}(E).
\]
The results for the approximations (\ref{eq:GF:2ab}) and (\ref{eq:GF:2ab:s-c}%
) are shown in Figs.\ \ref{fig:1}--\ref{fig:4}, and the critical values for
MIT are given in Table\ \ref{tab:1}. The numerical calculation were
performed for the square and two cubic lattices with a symmetric bare DOS.
We also treat the Bethe lattice, i.e. the model semielliptic bare conduction
band with
\[
N(E)=\frac 4{\pi W}\sqrt{1-\left( \frac{2E}W\right) ^2},
\]
($W$ is the bare bandwidth), the rectangular DOS, and the Gaussian DOS
\[
N(E)=\frac 4{\sqrt{2\pi }W}\exp \left( -2\left( \frac{2E}W\right) ^2\right)
\]
which corresponds to the hypercubic lattice in the large-$d$ limit. The
Gaussian DOS does not have band edges, so that the parameter $W$ is
determined from the second DOS moment,
\[
W=4\sqrt{\mu _2},\quad \mu _2=\int E^2N(E)dE.
\]
Then the quantity $\mu _2$ has equal expressions in terms of $W$ for the
Gaussian and semielliptic bare DOS's.

It is important that the quantity $F(E)$, unlike $F_0(E)$, does not diverge
at $E\rightarrow 0$, i.e. in the centre of the band. This fact is just due
to many-electron corrections. Therefore the non-self-consistent (NSC)
formulas (\ref{eq:GF:2ab}) yield a metal-insulator transition at $U\neq 0$,
unlike NSC local approximations. However, the corresponding critical value $%
U_{{\rm c}}^{{\rm NSC}}$ is rather small. The ``false'' singularities at the
edges of the Hubbard-I bands occur at large $U$ in our NSC approximation
(see the discussion in Ref.\ \cite{Anokhin:1991}).

The critical value for MIT $U_{{\rm c}}$ in the standard Hubbard-III
approximation \cite{Hubbard-III:1964} for an arbitrary bare DOS is given by
\cite{Anokhin:1991a}
\begin{equation}
U_{{\rm c}}^{{\rm H}}=2\sqrt{3\mu _2}.  \label{eq:Uc:HIII}
\end{equation}
The critical value in the SC approximation (\ref{eq:GF:2ab:s-c}) is changed
somewhat in comparison with the Hubbard-III result (see Table\ \ref{tab:1}).
Unlike Ref.\ \cite{Luo:2000}, where the critical value was decreased by
fluctuations, $U_{{\rm c}}/W=0.67$ for the Bethe lattice, our approach
yields an opposite tendency, in agreement with the results of the QMC
calculations at finite temperatures, $U_{{\rm c}}/W\simeq 1$ (see Table\ \ref
{tab:2}). Within the ``linearized'' DMFT \cite{Bulla:2000a}, an analytical
expression $U_{{\rm c}}$ can be obtained, which exceeds the Hubbard-III
value,
\begin{equation}
U_{{\rm c}}^{{\rm L}}=\sqrt{3}U_{{\rm c}}^{{\rm H}}=6\sqrt{\mu _2}.
\label{eq:Uc:LDMFT}
\end{equation}
As follows from comparison with Table \ref{tab:2}, this approximation seems
to overestimate somewhat $U_{{\rm c}}.$

The account of the Fermi excitations results in a modification of the DOS
form (cf. Ref.\ \cite{Anokhin:1991a}). In comparison with the Hubbard-III
approximation, a pronounced pseudogap exists near MIT at $U<U_{{\rm c}}$.
The same feature can be seen from the results of Ref.\ \cite{Luo:2000}. At
small $U$, a three-peak structure can be seen in Figs.\ \ref{fig:1}--\ref
{fig:4}, which becomes smeared with approaching MIT (the central peak
becomes wide, and a pseudogap occurs). The three-peak structure is more
pronounced in the NSC approximation. The details of our MIT scenario differ
from the DMFT picture where the central quasiparticle peak is expected to
shrink gradually at $U\rightarrow U_c-0$. Probably, this discrepance is
connected with the overestimation of the role of the damping in our
approach. The correct treatment of the damping is a difficult problem. To
avoid this problem, most large-$d$ calculations are performed at finite
temperatures. It should be noted that various DMFT versions give somewhat
different MIT pictures. In some calculations, a pseudogap develops in the
metallic phase near MIT (see, e.g., Refs.\cite{Jarrell:1992,Prushke:1993}),
as well as in our picture. The occurrence in some works of a very high and
narrow peak with lowering temperature, even not too close to $U_c,$ is
probaly an unphysical drawback. In a number of large-$d$ calculations (see
Refs.\cite{larged,Zhang:1993,larged1,Bulla:1999}) the height of the peak
does not change when approaching MIT.

In the case of square lattice, the situation is more complicated owing to
the Van Hove singularity at the band centre. The underestimation of $U_{{\rm %
c}}$ in the Hubbard-III approximation is confirmed by the calculations of
Ref.\cite{Mancini} where $U_{{\rm c}}\simeq 1.5W.$ More weak Van Hove
singularities are present for cubic lattices. One can see from Table 1 that
the difference between our calculations and Hubbard-III results becomes
rather strong for these lattices .

To conclude, we have demonstrated that a simple decoupling scheme enables
one to reproduce the non-trivial spectrum structure in the half-filled
Hubbard model. Our approach yields a qualitative agreement with the results
of large-$d$ approaches and QMC calculations. At the same time, this can be
easily applied for arbitrary two- and three-dimensional lattices. In
principle, the many-electron Hubbard operator method enables one to consider
in a regular way the problem of electron structure of systems with the
Hubbard splitting. Various types of slave boson and fermion representations
combined with diagram techniques can be used to this end.

Since our approach starts from Hubbard's subbands and includes large
incoherent contributions, this does not reproduce properly the Fermi-liquid
(FL) description of quasiparticle states. An account of low-energy spin and
charge dynamics would be useful to describe the electron spectrum picture in
more detail. A possibility of the transition from FL to non-FL behavior
which can take place near MIT should be also taken into account.

The work is supported in part by the Grant of RFFI No. 00-15-96544 (Support
of Scientific Schools).

\begin{table}[tbp]
\caption{Critical values of metal-insulator transition for different bare
DOS forms in the Hubbard-III approximation, $U_{{\rm c}}^{{\rm H}}$,
``linearized'' DMFT, $U_{{\rm c}}^{{\rm L}}$, and NSC and SC approximations (%
\ref{eq:GF:2ab}) and (\ref{eq:GF:2ab:s-c}), $U_{{\rm c}}^{{\rm NSC}}$ and $%
U_{{\rm c}}^{{\rm SC}}$.}
\label{tab:1}
\begin{tabular}{lcccc}
DOS & $U_{{\rm c}}^{{\rm H}}/W$ & $U_{{\rm c}}^{{\rm L}}/W$ & $U_{{\rm c}}^{%
{\rm NSC}}/W$ & $U_{{\rm c}}^{{\rm SC}}/W$ \\
\tableline rectangular & $1$ & $1.73$ & $0.99$ & $1.22$ \\
semielliptic & $\sqrt{3}/2=0.866$ & $1.5$ & $0.87$ & $1.06$ \\
Gaussian & $\sqrt{3}/2=0.866$ & $1.5$ & $0.87$ & $1.06$ \\
square & $0.866$ & $1.5$ & $0.87$ & $1.06$ \\
simple cubic & $0.707$ & $1.22$ & $0.76$ & $0.99$ \\
bcc & $0.612$ & $1.06$ & $0.67$ & $0.92$%
\end{tabular}
\end{table}

\begin{table}[tbp]
\caption{Critical values for the metal-insulator transition for the Bethe
lattice, $U_{{\rm c}}^{{\rm B}}$, and in the large-$d$ case, $U_{{\rm c}}^{%
{\rm G}}$, from different works.}
\label{tab:2}
\begin{tabular}{llll}
$U_{{\rm c}}^{{\rm B}}/W$ & $U_{{\rm c}}^{{\rm G}}/W$ & Refs. & Method \\
\tableline & $1.20$ & \cite{Jarrell:1992} & QMC \\
& $1.24$ & \cite{Prushke:1993} & QMC \\
$1.685$ &  & \cite{Zhang:1993} & PT\tablenote{Perturbation Theory.}, QMC \\
$1.262$ & $1.273$ & \cite{Georges:1993} & MFT, IPT\tablenote{Iterated
Perturbation Theory.} \\
& $1.273$ & \cite{Caffarel:1994} & QMC \\
$1.64$ &  & \cite{Rozenberg:1994} & MFT, QMC \\
$1.45$ &  & \cite{Moeller:1995} & PSCA\tablenote{Projective SC
Approximation.} \\
$1.2$ &  & \cite{Schlipf:1999} & QMC \\
$1.47$ & $1.45$ & \cite{Bulla:1999} & DMFT \\
$0.67$ &  & \cite{Luo:2000} & improved Hubbard III \\
$1.47$ &  & \cite{Bulla:2000} & NRG\tablenote{Numerical Renormalization
Group.}
\end{tabular}
\end{table}

\begin{references}
\bibitem[*]{E}  Electronic address: Valentin.Irkhin@imp.uran.ru

\bibitem{Hubbard-I:1963}  J. Hubbard, Proc. Roy. Soc.~A {\bf 276}, 238
(1963).

\bibitem{Hubbard-III:1964}  J. Hubbard, Proc. Roy. Soc.~A {\bf 281}, 401
(1964).

\bibitem{larged}  A. Georges and G. Kotliar, Phys. Rev.~B {\bf 45}, 6479
(1992); X. Y. Zhang and C. M. Zhang, Phys. Rev.~B {\bf 49}, 7929 (1994); F.
J. Ohkawa, J. Phys. Soc. Jpn. {\bf 61}, 1615 (1992).

\bibitem{larged1}  A. Georges, G. Kotliar, W. Krauth, and M. J. Rozenberg,
Rev. Mod. Phys. {\bf 68}, 13 (1996).

\bibitem{Rozenberg:1994}  M. J. Rozenberg, X. Y. Zhang, and G. Kotliar,
Phys. Rev. Lett. {\bf 69}, 1236 (1992); M. J. Rozenberg, G. Kotliar, and X.
Y. Zhang, Phys. Rev.~B {\bf 49}, 10181 (1994).

\bibitem{Schlipf:1999}  J.~Schlipf, M.~Jarrell, P.G.J.~van Dongen,
N.~Bl\"umer, S.~Kehrein, Th.~Pruschke, and D.~Vollhardt, Phys. Rev. Lett.
{\bf 82}, 4890 (1999).

\bibitem{Luo:2000}  H.-G. Luo and S.-J. Wang, Phys. Rev.~B {\bf 61}, 5158
(2000).

\bibitem{Anokhin:1991}  A. O. Anokhin and V. Yu. Irkhin, phys. stat.
sol.~(b) {\bf 165}, 129 (1991).

\bibitem{Anokhin:1991a}  A. O. Anokhin, V. Yu. Irkhin, and M. I. Katsnelson,
J. Phys.: Cond. Mat. {\bf 3}, 1475 (1991).

\bibitem{Irkhin:1999}  V. Yu. Irkhin and A. V. Zarubin, Eur. Phys. J.~B {\bf %
16}, 463 (2000).

\bibitem{Hubbard-IV:1965}  J. Hubbard, Proc. Roy. Soc.~A {\bf 285}, 542
(1965).

\bibitem{Irkhin:1994}  V. Yu. Irkhin and Yu. P. Irkhin, phys. stat. sol.~(b)
{\bf 183}, 9 (1994); cond-mat/9812072.

\bibitem{Bulla:2000a}  R. Bulla and M. Potthoff, Eur. Phys. J.~B {\bf 8},
565 (2000).

\bibitem{Jarrell:1992}  M. Jarrell and Th. Pruschke, Z. Phys.~B {\bf 90},
187 (1993).

\bibitem{Prushke:1993}  Th. Pruschke, D. L. Cox, and M. Jarrell, Phys.
Rev.~B {\bf 47}, 3553 (1993).

\bibitem{Zhang:1993}  X. Y. Zhang, M. J. Rozenberg, and G. Kotliar, Phys.
Rev. Lett. {\bf 70}, 1666 (1993).

\bibitem{Georges:1993}  A. Georges and W. Krauth, Phys. Rev.~B {\bf 48},
7167 (1993).

\bibitem{Caffarel:1994}  M. Caffarel and W. Krauth, Phys. Rev. Lett. {\bf 72}%
, 1545 (1994).

\bibitem{Moeller:1995}  G.~Moeller, Q.~Si, G.~Kotliar, M.~Rozenberg, and
D.S.~Fisher, Phys. Rev. Lett. {\bf 74}, 2082 (1995).

\bibitem{Bulla:1999}  R. Bulla, Phys. Rev. Lett. {\bf 83}, 136 (1999).

\bibitem{Bulla:2000}  R. Bulla, Adv. Solid State Phys. {\bf 40}, 169 (2000).

\bibitem{Mancini}  F.~Mancini, Europhys. Lett. {\bf 50}, 229 (2000).
\end{references}
\end{document}